\documentclass{aa}

\usepackage[varg]{txfonts}
\usepackage{graphicx}
\usepackage{upgreek}

\usepackage{ulem}

\begin{document}

\title{Correlations between laboratory line lists for FeH, CrH, and NiH and M-star spectra collected with ESPaDOnS and SPIRou\thanks{Our line lists are available in electronic format only. They can be obtained from the CDS via anonymous ftp to http://cdsarc.cds.unistra.fr/ or via https://cdsarc.cds.unistra.fr/  }}

\titlerunning{FeH \& CrH in  spectra from ESPaDOnS and SPIRou }

\author{P.~Crozet\inst{1} \and J.~Morin \inst{2} \and A.~J.~Ross\inst{1} 
\and S.~Bellotti\inst{3}$^,$\inst{4} \and J.~F.~Donati\inst{3} \and P.~Fouqu\'e\inst{3} \and C.~Moutou\inst{3}
\and P.~Petit\inst{3} \and A.~Carmona\inst{5}   \and A.~K\'osp\'al\inst{6}$^,$\inst{7}
\and A.~G.~Adam \inst{8} \and D.~W.~Tokaryk \inst{9}}
\institute{Universit\'e Claude Bernard Lyon 1 \& CNRS, Institute Lumi\`ere Mati\`ere, 69622 Villeurbanne, France
\and{Universit\'e de  Montpellier, CNRS, LUPM, 34095 Montpellier, France} 
\and {Universit\'e de Toulouse   CNRS, IRAP, 14 av. Belin, 31400 Toulouse, France}
\and  {Science Division, Directorate of Science, European Space Research and Technology Centre (ESA/ESTEC),
             Keplerlaan 1, 2201 AZ, Noordwijk, The Netherlands} 
\and{Universit\'e Grenoble Alpes \& CNRS, IPAG, 38000 Grenoble, France}             
\and {Konkoly Observatory, HUN-REN Research Centre for Astronomy and Earth Sciences (CSFK, MTA Centre of Excellence),  Konkoly-Thege Mikl\'os \'ut 15-17, 1121 Budapest, Hungary}
\and {Max Planck Institute for Astronomy, K\"onigstuhl 17, 69117 Heidelberg, Germany }
\and {Department of Chemistry, University of New Brunswick,  Fredericton, NB, Canada E3B 5A3}
\and {Department of Physics, University of New Brunswick,  Fredericton, NB, Canada E3B 5A3}}

\date{Received    18/07/2023
Accepted  19/09/2023  }

\abstract {Molecular bands of metal oxides and hydrides dominate the optical and near-infrared spectra of M dwarfs. High-resolution spectra of these bands have immense potential for determining many properties of these stars, such as effective temperature, surface gravity, elemental abundances, radial velocity, or surface magnetic fields. Techniques are being developed to do this but remain limited by the current availability and accuracy of molecular data and spectral line lists.} 
{This paper reports metal monohydride line lists selected from near-infrared and visible laboratory data to show that specific bands in several electronic transitions can be used to identify CrH, NiH, and FeH in M stars and to determine radial velocities from Doppler shifts. 
The possibility of measuring magnetic fields is also investigated for FeH and CrH.}  {We used systematic cross-correlation analysis between unpolarised spectra from a selection of M stars and state-specific laboratory line lists. These lists were generated from a combination of existing data and new laboratory laser-excitation spectra recorded at Doppler-limited resolution, in zero-field conditions or in magnetic fields up to 0.6 tesla.} {We show that transitions at visible wavelengths in FeH and NiH, usually neglected in the analysis of the spectra of M-type stars, do in fact contribute to the spectra, and we demonstrate the influence of magnetic sensitivity on selected transitions in CrH and FeH.}
 {Although the new line lists focus on transitions recorded at temperatures significantly lower than those of stellar objects, they remain pertinent because they cover some band-head regions of high spectral density.  FeH bands can provide a useful supplement to atomic lines for the analysis of high-resolution optical and near-infrared spectra of M dwarfs. We demonstrate the influence of a magnetic field on CrH signatures around 862 nm.}


\maketitle

\section{Introduction}
The extended study of cool stellar objects, and concurrent discoveries of exoplanets, has led to a demand for more reference data of diatomic molecules seen in stellar atmospheres, species that can also be formed in plasma and flame environments in the laboratory. 
Accurate line lists, that include as many relevant species and transitions as possible,  are essential for generating synthetic spectra to improve determinations 
of stellar properties, such as temperature, Doppler shifts, stellar magnetic fields, and rotation, either from direct comparison or from cross-correlation (CC) calculations. 
Precise and complete line lists -- including molecules -- for M dwarfs will be particularly valuable if such data can explain the diverse behaviour of several subsets of spectral lines, with key applications for 
velocimetry and magnetometry \cite[e.g.][]{Klein2021, Bellotti2022}.  We consider here some molecular transitions expected to be sensitive to magnetic fields, aiming to identify and characterise
 those likely to have sharp and deep features that should be well suited for such diagnostics via CC calculations.  
 Our laboratory work on FeH, NiH, and CrH falls in the visible and near-infrared regions; we present comparisons with  data
 from  ESPaDOnS  (Echelle SpectroPolarimetric Device for the Observation of Stars; \citealt{Donati2003}),  which operates in the optical region 0.37--1.05 $\upmu$m, its twin instrument Narval \citep{Auriere2003} 
 at the Pic du Midi, and SPIRou  (SpectroPolarim\`etre InfraRouge; \citealt{Donati2020}) in the near-infrared (0.95--2.5 $\upmu$m). These spectropolarimetric data have already been analysed 
 in the context of stellar magnetism  and radial velocity (RV) determinations \citep{Donati2008, Morin2008, Morin2010, Cristofari2023, Carmona2023, Bellotti2023}.
 
Cross-correlation is a powerful tool that has been successfully applied in a variety of scientific fields, from signal processing and wireless communications to quantum physics.
 In the context of stellar astrophysics, Doppler shifts supply one correlation function linking stellar spectra with laboratory-measured `masks', 
 consisting of reference collections of atomic or molecular transitions. Information can be extracted at several levels, starting with whether the species is present. Spectral details then lead to 
estimates of temperature (at thermal equilibrium or otherwise), RV, magnetic field characteristics, surface gravity (through pressure broadening of spectral lines), and relative chemical abundances.
  Precise RV determinations  have been used to detect exoplanets around stars for more than 2 decades, following evidence from \citet{Mayor_1995} of an extra-solar planet around a Sun-like star. 
Their CC approach was introduced for cool dwarfs of spectral type FGK that feature mostly atomic lines and a well-defined continuum. Moving towards cooler  M dwarfs, atomic 
 transitions become fewer and weaker, and molecular bands emerge. Molecular line lists are notoriously less precise than atomic ones, and purely synthetic masks can be 
 insufficiently accurate to give precise RVs. Different approaches  have been proposed to 
 compensate for these issues. Optical developments and new algorithms to describe the instrumental response of the High Accuracy Radial velocity Planet Searcher (HARPS) spectrometer have resulted in precise 
 RVs via `template matching', effectively averaging many spectra of the object of interest \citep{Anglada2012}. An alternative `line-by-line'  approach, employed for example in papers by \citet{Dumusque2018}, \citet{Lafarga2020},  and \citet{Artigau2022},
 determines RVs from individual features,  then recognises and excludes outliers prior to averaging the outcomes.  
Molecular templates have also proven helpful in probing exoplanetary atmospheres.  Pioneering work by \citet{Snellen2010} extracted evidence for strong winds on exoplanet HD209458b from RV results  derived from CC functions with resolved CO lines around 2.3 $\upmu$m. The review by \citet{Birkby2018} illustrates the importance of information from high-resolution (ground-based) instruments for the detection and characterisation of exoplanets and their atmospheres, and highlights the power of the CC approach. 
  
Absorption spectra from cool stellar objects display rovibrational and/or electronic bands of small molecules (mostly diatomic oxides and hydrides, and water) as well as strong atomic features, 
  providing further insight into chemistry and conditions.  There is some bias induced by the intrinsic strength of the optical transitions, with the lighter Group II 
  and first-row transition metal diatomics dominating the picture. Their numerous molecular signatures, particularly in the near-infrared (Y, J, H, and K  bands) allow the
   identification of species in photospheres with certainty, thanks to precise molecular line lists, typically with $\Delta\uplambda$/$\uplambda \leq 2\times10^{-7}$, widely distributed in rotational 
   quantum numbers.  Because many of the species stable above 1000 K are open-shell fragments of chemically stable laboratory precursors, many of them 
   are sensitive to magnetic fields. Magnetic response depends on the changes in quantum numbers (rotation and electronic) associated with a given spectral line, 
   and this results in some transitions being effectively insensitive to the magnetic field whilst others show pronounced Zeeman effects.  It is helpful to select the appropriate 
   subset of reference lines to infer the influence of magnetic fields, assuming the relevant Land\'e factors are known. Attempts to model the Land\'e factors
   for transition metal monohydride radicals, either from laboratory work \citep{Harker2013,Crozet2014} or from solar and stellar spectra \citep{Afram2008,Shulyak2010},
   show that they are not trivial to predict. 

Of course, there are limitations and complications. Contributions from different systems frequently overlap one another: TiO bands \citep{McKemmish2019} stretch across 
   the visible and near-infrared spectrum, and electronic bands of atmospheric oxygen overlap part of the A-X system of CrH at 764 nm. Whenever possible,
   it is preferable to work in a spectral range with a single dominant contribution, such as FeH around 1 micron \citep{Wende2010}.    It is also true that Land\'e factors 
   have been measured for few of the molecules of interest except at very low temperatures, notably from molecular beam experiments conducted in Steimle's group \citep{Virgo2005,Chen2006,Chen2007,Harrison2008,Qin2012} for TiO, TiH, CrH, and CaH, for example. 
   Work is in progress  in Lyon and in Fredericton, at the University of New Brunswick (UNB), to supply this information for selected transition metal hydrides, using high-resolution laser spectroscopic techniques.  

We collated metal monohydride (MH) transition wavenumbers from our own laboratory work and from literature sources to target reasonably narrow spectral windows across the near-infrared and visible regions.  
  We then used them to identify MH transitions in some M-star spectra recorded at telescopes at Pic du Midi (Bernard Lyot Telescope) and on Maunakea (Canada-France-Hawaii Telescope), 
  and to extract RVs via CC calculations.  The values from the molecular data are found to match values derived from atomic masks,   even though it is sometimes difficult to recognise spectral features at a glance.  
  The band-heads of the  A~$^6\Sigma^{+}$ - X~$^6\Sigma^{+}$ system of CrH and F~$^4\Delta$ - X$^4\Delta$ system of FeH emerge clearly in the stellar spectra we have selected, but
  NiH contributions do not. They are spread across the visible region and overlap with strong contributions from TiO.  Even though individual lines cannot be identified, we find that NiH is present in some sources. 
   It seems to be a genuine result because a `deuterium test', following \citealt{Pavlenko2008} and performing the same calculation using  a mask constructed for NiD (not expected to be present),  gives a negative outcome.  
   CH, SiH, and NH radicals are already included in the widely used  Vienna Atomic Line Database (VALD) \citep{Ryabchikova2015},  but only one metal hydride, MgH, is found there\footnote{https://www.astro.uu.se/valdwiki/Molecules}.  We make a case for adding a few more. 
  Sources such as MoLList \citep{Bernath2020} and ExoMol  \citep{Tennyson2020} offer more extensive, high-temperature line lists for FeH and CrH, but not all entries match experimental precision.  
  This  contribution is intended as a supplement and not a replacement, since the focus of our work  is on defining selective diagnostic transitions.
     
In the following, we give a description of the CC approach used here, to be applied with narrow-band spectral masks, and then discuss some optical transitions 
suitable as molecular templates in the investigation of spectra from some M-dwarf stars, recorded on the ESPaDOnS, Narval, and SPIRou spectropolarimeters. We use CC to show
that these systems retrieve RVs with degraded signal-to-noise ratios (S/Ns) for magnetically active objects. We use CrH to illustrate that the density of the spectrum 
in the presence of a magnetic field is such that very few sharp, deep features remain in the mask. 
The RV determined in this way is closer to reference data, but the statistical uncertainty remains large.

\section{Cross-correlation functions}
The CC method described by \citet{Tonry_Davis_1979} and by  \citet{Queloz1995} seeks the optimum match between a stellar spectrum and a reference mask, when the mask is Doppler-shifted by small steps of RV. We took  sampling steps of 0.5 km\,s$^{-1}$ in our calculations.
The correlation curve should ideally be a flat baseline with a single sharp peak. The mask can be either a simulated model-star stick-spectrum or a fully calibrated line list obtained in the laboratory. 

Two strategies emerged for the intensity of the reference mask: the elements of the stick spectrum can either be weighted, assuming relative intensities are known, or given a uniform influence (binary mask).  
 \citet{Pepe2002} showed that significant improvement in the S/N of the correlation curve can be achieved with proper weighting, but intensity data are not always available. 
We investigated the performance of weighted and binary masks on a subset of data for FeH where lines were well resolved, and found very little difference in the S/N, or the RV result. 
A more sophisticated approach would incorporate weighting according to the S/N of the stellar spectrum. We have not done this, as our tests are performed over small spectral regions, but 
 this strategy would have to be implemented were CC functions to be calculated over a very wide domain, with different line parameters and S/Ns.

If the correlation calculation gives a successful outcome, the peak is fitted to a Gaussian function, yielding a RV in the heliocentric rest frame
(assuming the barycentric correction has already been applied to the telescope input data). A S/N can be obtained by taking
 the amplitude of the correlation curve lying outside the correlation peak as noise.  We developed procedures to do this in Igor pro software (Wavemetrics Inc).  
\section{Telescope spectropolarimetric data and laboratory line lists}
\subsection{Stellar spectra from ESPaDOnS, Narval, and SPIRou}
Because of our interest in the Zeeman effect in MH molecular radicals, we selected data for some M-type dwarfs that span a wide range of magnetic activity, including several stars that are known to host very strong magnetic fields as test cases. 
 Spectra recorded on ESPaDOnS and its twin instrument Narval came from the PolarBase \citep{Petit2014} data archive of stellar spectropolarimetric observations. 
 Spectra from SPIRou  were measured as part of the SPIRou Legacy Survey\footnote{https://SPIRou-legacy.irap.omp.eu} described by \citet{Donati2020}, and processed as part of the follow-up SPICE programme\footnote{SPICE: SPIRou Legacy Survey - Consolidation and Enhancement, with observation time 2022-2024}. 
 SPIRou wavelength-calibrated telluric-corrected spectra and polarimetry data products were obtained using the APERO pipeline \citep{Cook2022} installed at the SPIRou Data Centre hosted at the Laboratoire d'Astrophysique de Marseille. 
 The spectrum for GJ~873 (EV Lac) is a template constructed by averaging 706 spectra recorded between 24 May 2018 and 13 March 2022,  with  a S/N >~72. The template for GJ~388 (AD Leo) averaged 348 spectra recorded  between 24 May 2018 and 3 November 2020, with  S/N >~86.  The template for GJ~411 averaged 947 spectra taken between 26 April 2018 and 12 May 2023.
 Table~\ref{table:stars} lists some reference average magnetic field strengths, <B>, established from Zeeman-broadening and intensification analysis \citep{Shulyak2017,Cristofari2023}, of (mostly) Ti~\textsc{i}, Fe~\textsc{i,} and FeH signatures.

\begin{table}
\caption{Identification of the cool stars chosen for this study, their average magnetic
field, <B>, and their RVs with respect to the heliocentric rest frame. Uncertainties are indicated in parentheses, in units of last digit.
<B> values are taken from publications by \citet{Reiners2022}$^a$, \citet{Kochukhov2021}$^b$, \citet{Shulyak2017}$^c$, and from \citet{Cristofari2023}$^d$.  RV values come from supplementary data associated with work by \citet{Lafarga2020}.}
         
\label{table:stars}  
 \centering                           
 \resizebox{\columnwidth}{!}{%
\begin{tabular}{l l l l l}          

\hline                     
~Star  & Instrument & <B>  & RV  \\  
&  & Tesla &      km\,s$^{-1}$   & \\  
\hline                                   
\\
GJ~411                  & SPIRou                & <0.02$^a$             & -85.02 (2) \\
GJ~1002                         & ESPaDOnS      &  0.08$^b $            & -40.12(2)  \\       
GJ~49                           & Narval                & 0.08((2)$^c$          & -6.30 (2)   \\
GJ~406 (CN Leo)         & ESPaDOnS      & 0.345(20)$^d$         & +19.28 (2)  \\
GJ~388 (AD Leo)         & SPIRou                & 0.303(6)$^d$          & +12.29 (2)\\
GJ~873 (EV Lac)         & SPIRou                & 0.453(7)$^d$          & +0.35 (3) \\
GJ~412B (WX UMa) & ESPaDOnS     & 0.73(3)$^c$           & +69.16(2)  \\ 
\hline                                   
\end{tabular}}
\end{table}
 
\subsection{Laboratory line lists}
\begin{figure}[!h]
  \centering\includegraphics[scale=0.25]{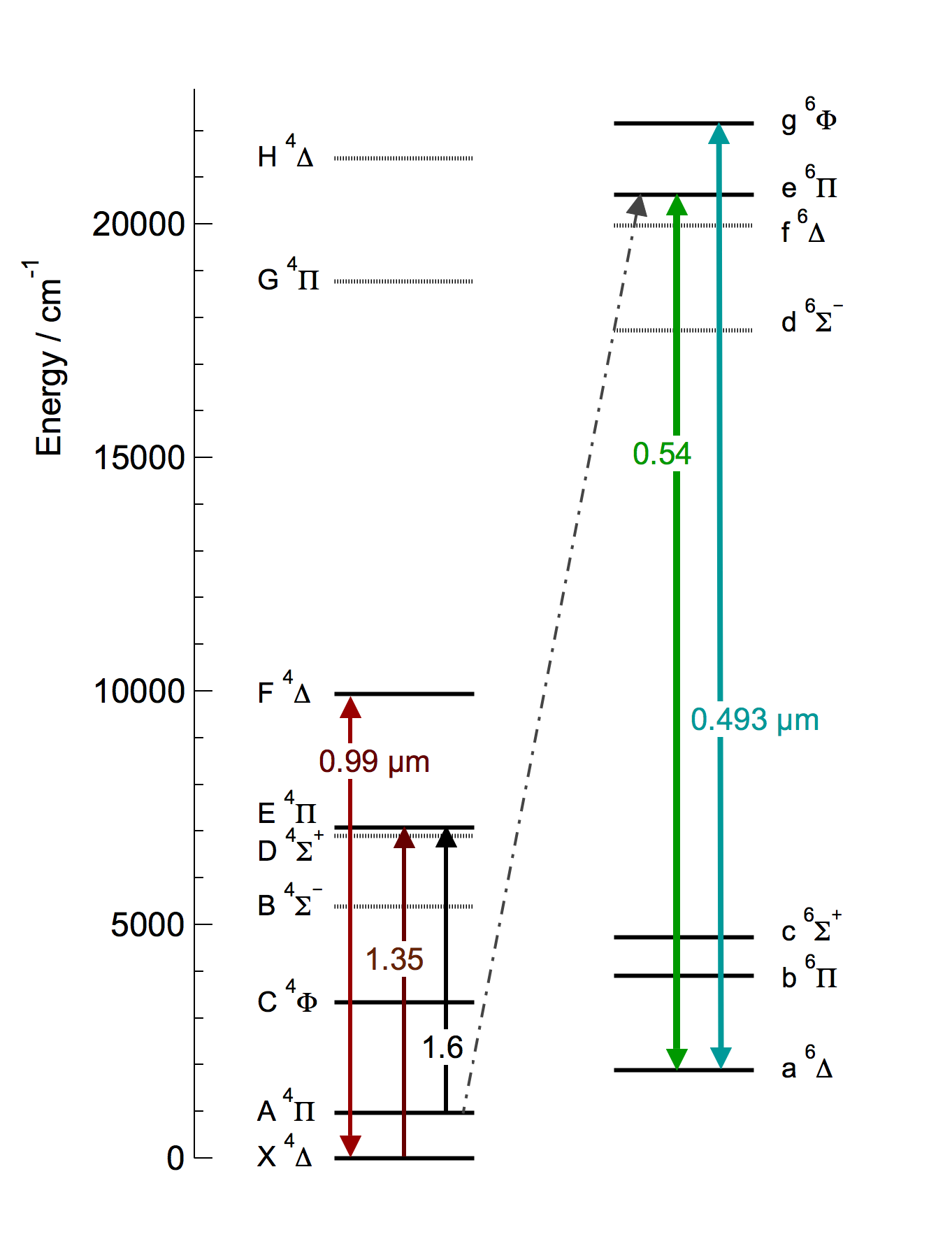}
 \caption{Some quartet and sextet electronic states of FeH, following \citet{Wilson2001}. Arrows denote observations as excitation or fluorescence. 
 Solid lines show the energy of the lowest spin-orbit component for observed electronic levels. Faint, broken lines 
 indicate calculated (but currently unreported) levels from \citet{Langhoff_1990}.}
  \label{fig:FeHAbInitio}
\end{figure}

\paragraph{FeH data in the visible.} Most of the high-resolution spectroscopy of FeH in the visible region was performed by the J.M. Brown group in Oxford in the 1990s, reacting Fe(CO)$_5$ with atomic hydrogen
 \citep{Fletcher_1990,Carter_1993,Goodridge_1997,Wilson2001,Wilson_Brown2001}. 
 Several electronic transitions are known, as illustrated in Fig.~\ref{fig:FeHAbInitio}, originating from the electronic ground state, and from low-lying sextet states. 
 Strong transitions were observed from the lowest (metastable) sextet state, a~$^6\Delta$, which was populated under their laboratory conditions.  
 $\Delta\Omega$ =-1 and 0 transitions in the e~$^6\Pi$  - a~$^6\Delta$ system around 535 - 540 nm \citep{Goodridge_1997,WilsonCook2001,Wilson2001}, 
 $\Delta\Omega$ = 0 and +1 transitions in the  g~$^6\Phi$  - a~$^6\Delta$ system near 493 nm \citep{Carter1994} have been reported in the literature. 
 We took information on the origin (0-0) bands of both systems to create masks for CC calculations.  
 The e~$^6\Pi$  - A~$^4\Pi$ transition at 510 nm indicated with dash-dot line in Fig.~\ref{fig:FeHAbInitio} is formally spin-forbidden and weak. 
 It has not been included in the compilation for CC, but will provide a route to experimental 
 determination of Land\'e factors for the low-lying A~$^4\Pi$ state in future work. 
 
\paragraph{FeH data in the near-infrared.}
  Two quartet systems are well-known contributors to spectra of cool stellar objects  and sunspots. 
 The F~${^4\Delta}$ - X~$^4\Delta$ `Wing-Ford' band near 0.99 $\upmu$m studied in detail for example by \citet{Wende2010} and by \citet{Shulyak2019} is accessible to ESPaDOnS, Narval and SPIRou, although it falls right at the edge of SPIRou's spectral response. The 
  E~${^4\Pi}$ - A~$^4\Pi$  bands around 1.6 $\upmu$m investigated by \citet{Wallace2000}) and  E~${^4\Pi}$ - X~$^4\Delta$  bands around 1.35 $\upmu$m 
  are much better placed for the optics of the SPIRou instrument. We took line lists from the Kitt Peak National Solar Observatory (NSO) database 
 \citep{Balfour2004,Hargreaves2010} to apply the CC method to the spectra of some M stars recorded by SPIRou.
  
\paragraph{CrH data.} 
   The origin (0-0) band of the A~$^6\Sigma^{+}$ - X~$^6\Sigma^{+}$  system of chromium monohydride near 862 nm is a marker proposed by \citet{Kirkpatrick_1999} and by \citet{Burrows2002}
 for L-type brown dwarfs.  Slightly weaker, the (1-0) band near 764 nm is conveniently placed for laboratory studies, but it overlaps with the 
 atmospheric A-band of O$_2$, and is therefore less useful for astrophysical purposes. The  A-X (0-1) CrH band at 996.9 nm  mentioned by \citet{Reiners2007} 
 is also quite strong, but lies beneath the even stronger (0-0) Wing-Ford band of FeH. Still further to the infrared,  in the  J-band  around $\sim1.2~\upmu$m, weaker bands of FeH and CrH
  again appear together, although the CrH band-head falls in a  sparser region of FeH, according to simulations by \citet{Dulick2003}.  
 
 \citet{Kuzmychov2013}  noted that CrH could provide a very sensitive diagnostic for magnetic field, 
 with the degeneracy of rotational levels splitting very quickly in the presence of very modest magnetic fields. The Zeeman/Paschen-Back effect in the electronic ground state has 
 been studied and modelled in the laboratory \citep{Corkery_1991,Lipus_1991}, but the excited states are much more challenging, with very strong perturbations 
 making the spectra very irregular even in zero-field conditions, as noted by several authors, notably \citet{Ram_1993}, \citet{Bauschlicher2001}, and \citet{Chowdhury2006}. 
 
\begin{figure}[!h]
 \centering\includegraphics[scale=0.25]{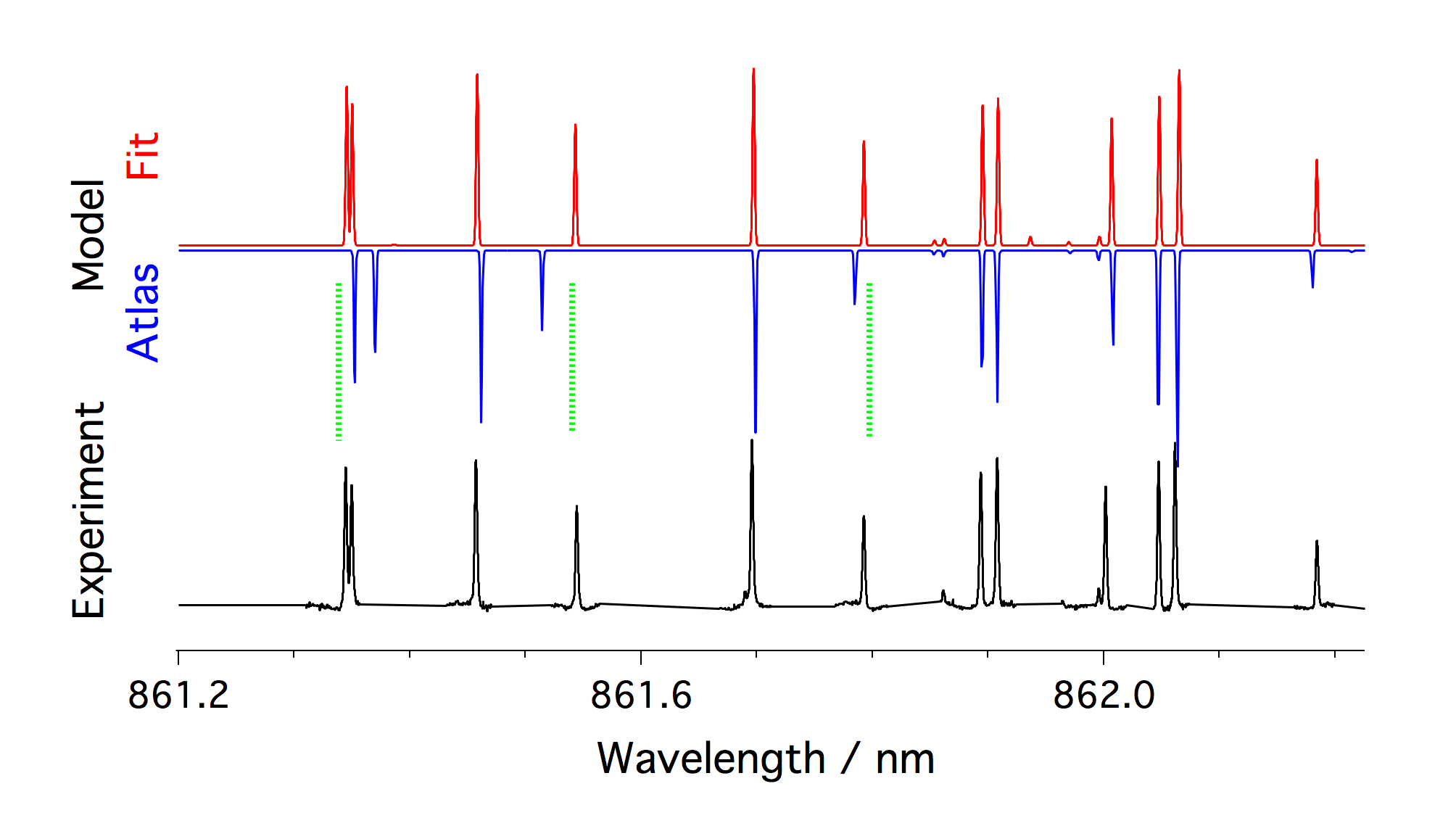}
\caption{R-head region of the 0-0 band  of $^{52}$CrH A-X, at arbitrary scale. Shown in red is our fit, N' $<$ 15.\ In blue is the calculated absorption cross-sections from the ExoMol software \citep{Tennyson2020} based on the data from \citet{Burrows2002};  the
full intensity scale for the strongest line shown is 8 $\times$ 10$^{-15}$ cm${^2}$/molecule. 
A laboratory spectrum is shown in black. Dashed green lines indicate significant  improvements in our model.}
\label{fig:CrHLifvsExomol}
\end{figure}

  The best available CrH atlas, from \citet{Burrows2002} combines constants derived from unperturbed transitions with ab initio results (including transition dipole moments) and covers 
 the electronic spectrum of CrH from 0.67 to 1.0 $\upmu$m. This atlas is part of the \cite{Bernath2020} MoLLIST database. It has been adopted on the ExoMol website developed by \citet{Tennyson2020}, 
 and can be used to generate a synthetic spectrum at any desired temperature or resolution. This work nevertheless fails to reproduce the true positions of many spectral features, 
 with `observed' minus `atlas' wavenumbers sometimes in excess of 1 cm$^{-1}$;  this was reported as a handicap for applications in stellar spectroscopy by \citet{Pavlenko2005}  and \citet{Reiners2007}. 
  Focusing on the 0-0 band-head regions,  the Lyon group re-recorded laser excitation spectra of CrH produced in a discharge source in order to fine-tune the reference line list. 
  An `effective parameter' fit with the PGOPHER program from \cite{Western2017} can reproduce our transitions with rotational quantum number N$^{\prime}\le15$ with an rms deviation < 0.025 cm$^{-1}$, 
  assuming an interaction between a single vibrational level of a ${^4\Sigma^+}$ state and v=0 of A~${^6\Sigma^+}$.  Figure \ref{fig:CrHLifvsExomol} illustrates the situation, with the upper traces corresponding to 
  the PGOPHER and Atlas models, and the lower trace a small part (11598--11612 cm$^{-1}$) of a laboratory laser-excitation spectrum at Doppler resolution. 
  We also recorded the R-head of the (0-0) band of the A-X system of CrH when forming the molecule between
 permanent magnets, and established that PGOPHER reproduces the observed Paschen-Back splittings properly for fields up to 0.5 tesla.
 A PGOPHER data file is available in electronic format from the CDS, as supplementary material.   

 Although a few inaccurate transition wavenumbers will not prevent CrH from being recognised in a spectrum, they will be critical for magnetic field diagnostics near the band origin, 
 where the lowest rotational lines produce the best-resolved magnetic components, and the most distinctive patterns in Stokes-V signals.
 
\paragraph{NiH data.}
 The low-lying coupled states of NiH have been studied by Fourier-transform resolved laser-induced fluorescence, with a view to understanding the rovibronic structure of this relatively simple 
system (just three electronic states are close in energy, each with a single unpaired electron \citep{Gray_1991,Abbasi2018,Havalyova2021}. 
 The transitions measured during this project now form a stand-alone low-temperature line list, and we used this for the CC calculations here. 
 The original dataset from \citet{Vallon2009} has since been extended further to the blue \citep{Ross2012,Harker2013}; the corresponding data for $^{58}$NiH 
  are available on the ExoMol database\footnote{https://www.exomol.com}.  \citet{Ross2012} and \citet{Harker2013} also recorded and analysed  Zeeman patterns for many but not all of these transitions. 

\section{Results and discussion}
\subsection{FeH sextet line lists (visible)}
\begin{figure}[!h]
 \centering\includegraphics[scale=0.25]{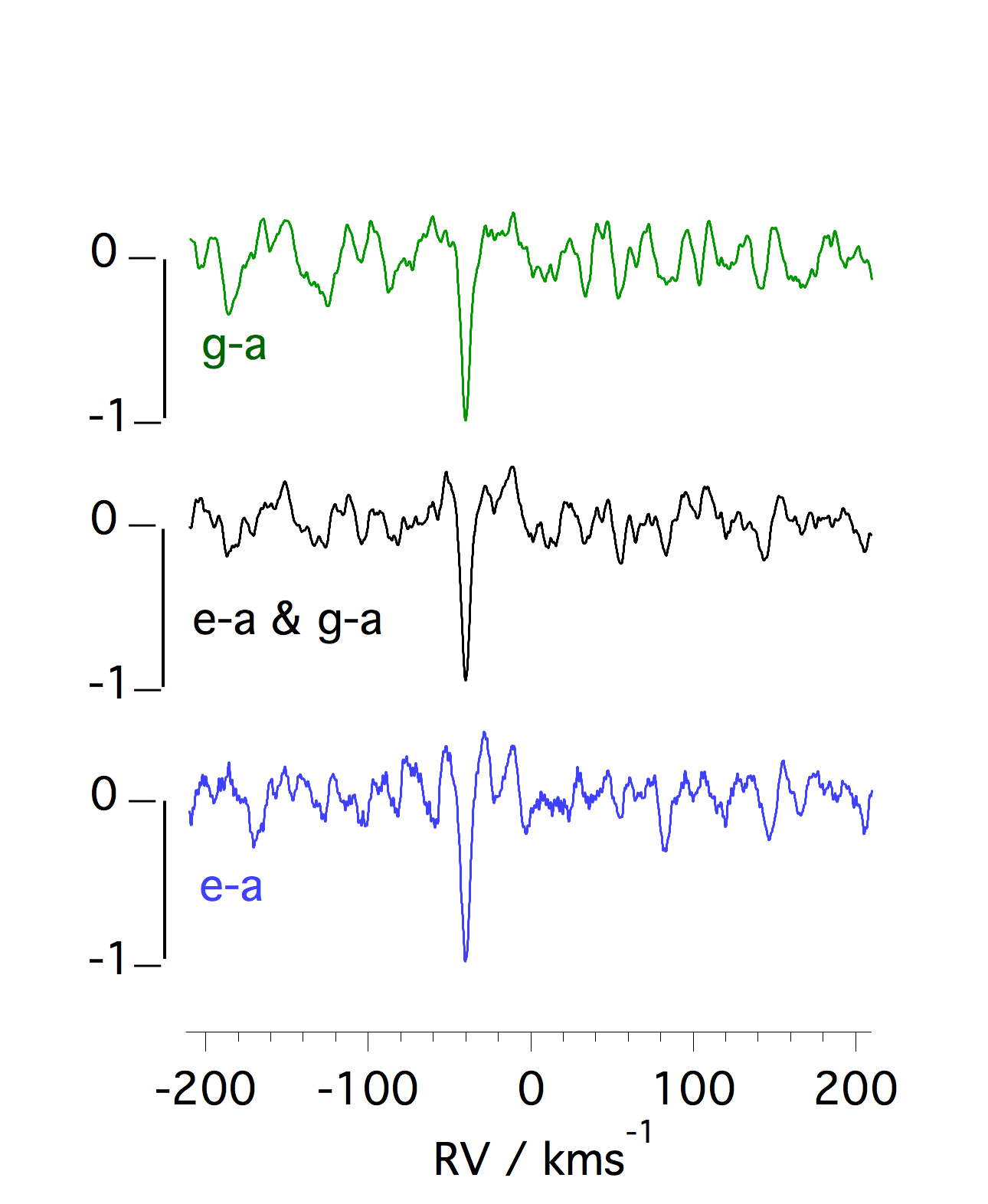}
\caption{CC functions between the Stokes-I spectrum of the weakly magnetic star GJ~1002 and (g-a), (e-a), and (g-a + e-a) masks. 
The curves are shifted vertically for convenience. The S/Ns on correlation peaks are 7.8 from (g-a), 7.5 from (e-a), and 10.6 from the combined mask.}
\label{fig:FeHCCeaga}
\end{figure}

 The results of three CC calculations for FeH are shown in Fig.~\ref{fig:FeHCCeaga}. The uppermost trace shows the CC function using the g~$^6\Phi$ ~-~a~$^6\Delta$ mask alone 
(117 reference lines in range 20058-20459 cm$^{-1}$, or 488.65-498.41 nm), and the lowest one uses 116 $\Delta\Omega=\Delta\Lambda$ lines in range 18621-18952 cm$^{-1}$ (527.50-536.88 nm) 
in the e~$^6\Pi$~-~a~$^6\Delta$ system.  The RVs determined from Gaussian profile fits to the  e~-~a and g~-~a  correlation peaks differ by 75 m/s. 
The central trace shows the sharper and cleaner result obtained when combining these two. 

\begin{figure}[!h]
 \centering\includegraphics[scale=0.28]{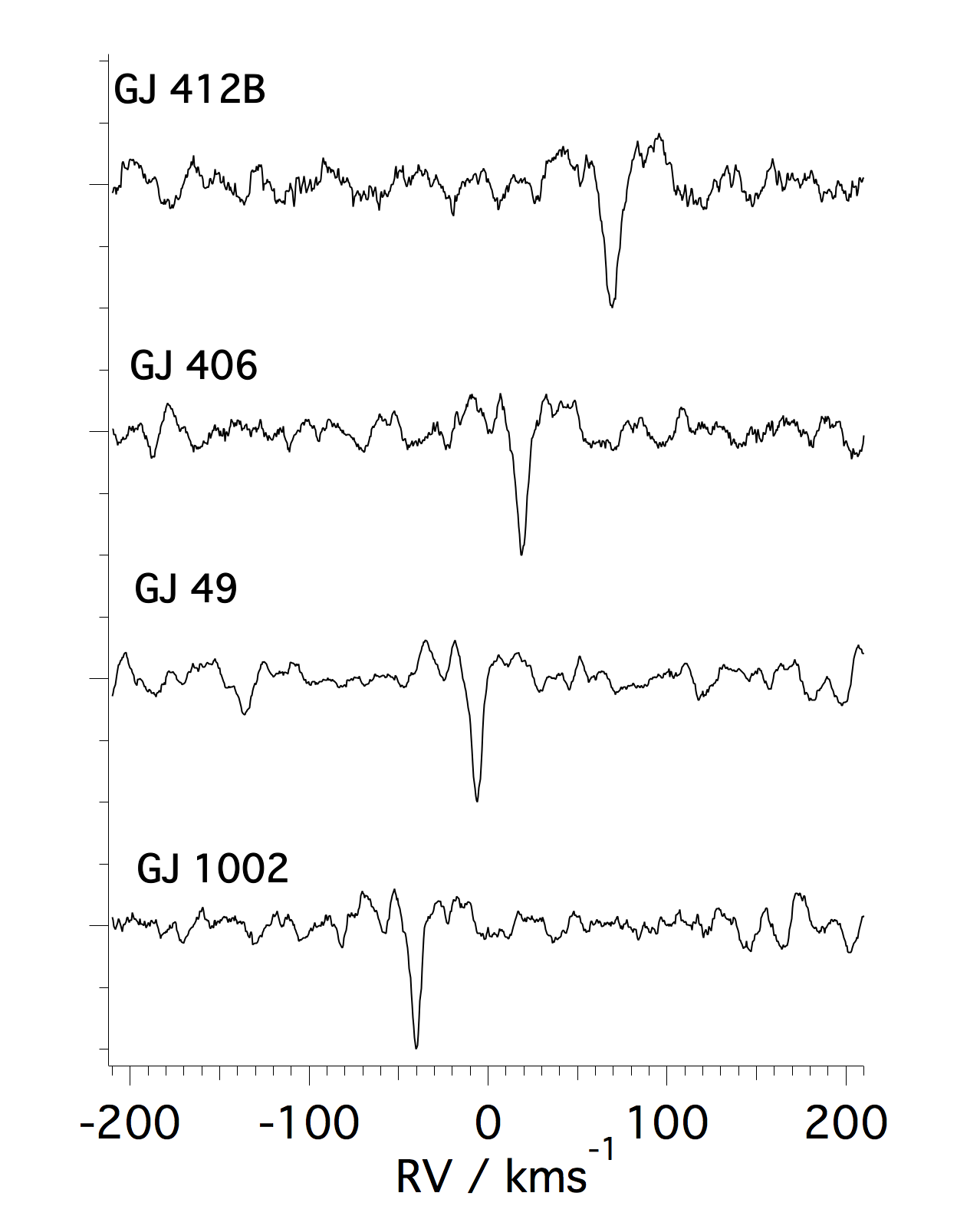}
\caption{CC curves (normalised to unity) for Stokes-I spectra of the four sample M stars using the combined (g-a plus e-a) mask, with 213 lines.  
Fits to Gaussian profiles gave:  GJ~412 B : RV= 69.1(2), fwhm = 8.8(4) km\,s$^{-1}$; GJ~406 :  RV= 19.45(3), fwhm = 6.9(4) km\,s$^{-1}$;
 GJ~49  RV= -6.4(1), fwhm = 6.7(3)   km\,s$^{-1}$; and GJ~1002 : RV= -40.01(3), fwhm = 5.9(3) km\,s$^{-1}$. The 1$\sigma$ uncertainty is quoted in parentheses in units of last digit.}
 \label{fig:FeHRVpeaks}
\end{figure}

The g-a and e-a line lists give consistent RV determinations for GJ~1002, (RV=-40.01$\pm$ 0.03 km\,s$^{-1}$), in good agreement with the literature value indicated in Table~\ref{table:stars}. 
  Combining the two masks results in a S/N improvement of the order of $\sqrt2 $ in the central trace of Fig.~\ref{fig:FeHCCeaga}, as expected from approximately doubling the number of lines in the mask.  
 We retained the combined (g-a plus e-a) mask, and applied it next to spectra from increasingly magnetically active stars. The results are shown in Fig.~\ref{fig:FeHRVpeaks}, 
  with CC functions for four of the stars listed in Table~\ref{table:stars}.  
  All four show sharp correlation peaks, confirming that these sextet transitions do contribute to the M-star spectra. The S/Ns between 7 and 10 in the correlation peaks suggest that this contribution is significant. 
  The RVs reported in Table~\ref{table:RVresults}  are very close to those obtained from the literature, from \cite{Lafarga2020}. 
  The 30\% increase in full-width at half maximum from lowest to topmost trace in 
  Fig.~\ref{fig:FeHRVpeaks} probably reflects a combination of increasing stellar rotation and magnetic activity.
To ensure that these correlation peaks are not chance matches, we ran the same CC calculations with line lists for FeD reported recently by \cite{Harvey2023}. 
As expected, none of the stellar spectra produced a correlation peak with FeD data.

\begin{figure} [!h]
 \centering\includegraphics[scale=0.20]{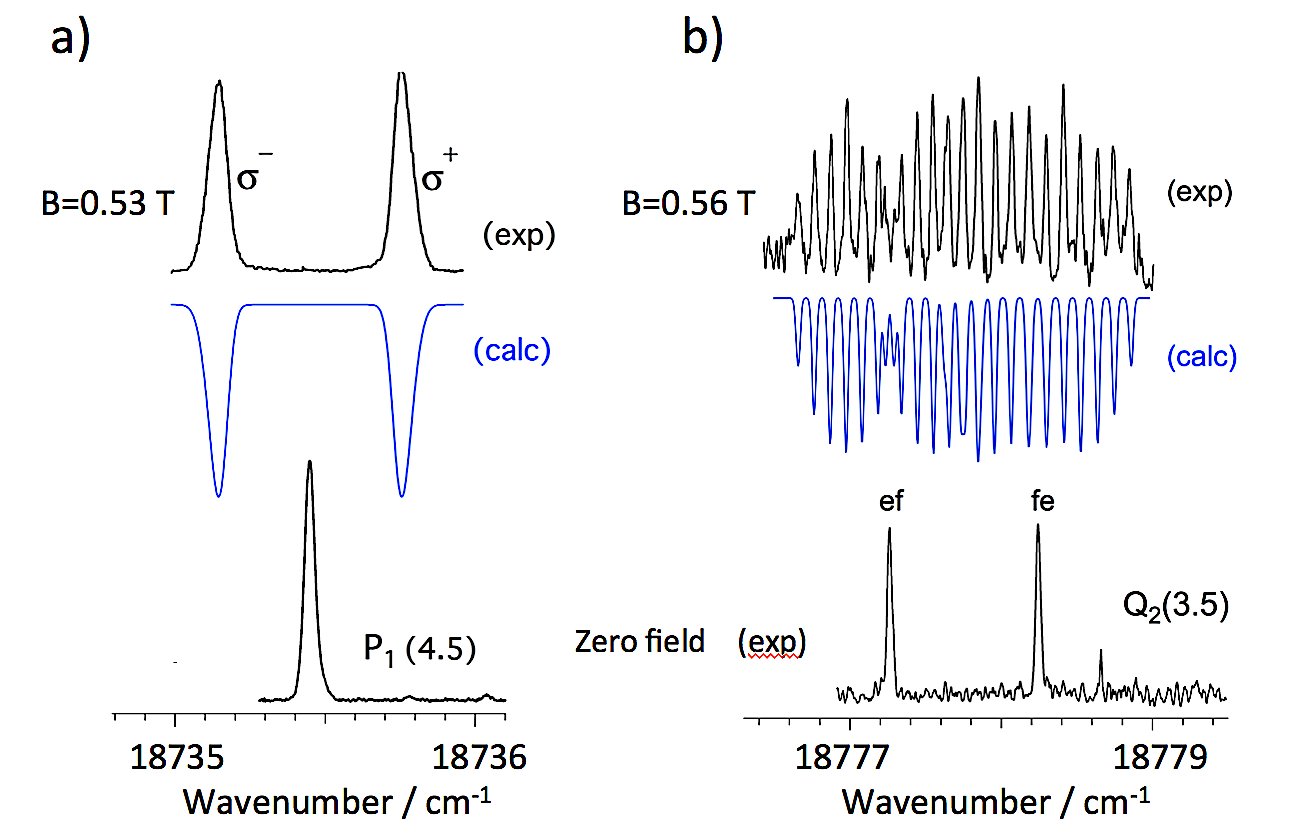}
\caption{Zeeman patterns in the e-a system of FeH, showing (a) the P$_1$(4.5) line ($\uplambda_{\textnormal{vac}}$=533.76 nm) at zero field and at  0.53 T, with unresolved $\Delta$M$_J$ =  +1 and -1 components equidistant from the zero-field line centre, and 
(b) the Q$_2$(3.5) line ($\uplambda_{\textnormal{vac}}$=532.54 nm), recorded at 0.56 T, with better-resolved structure. Note the extended wavenumber scale.  Simulations (calc) use effective g$_J$ Land\'e factors optimised from the experimental profiles.}
 \label{fig:FeHeaZeeman}
\end{figure}

\paragraph {Laboratory results.}
The question then becomes how to generate reliable effective Land\'e factors for each of the transitions listed in the g~$^6\Phi$~-a~$^6\Delta$ and e~$^6\Pi$  - a~$^6\Delta$ datasets. 
 Our laboratory work establishes that the electronic states involved do not follow pure Hund's coupling cases (a) or (b) according to the state symmetry, so some 
 work is required to guide models, before the least squares deconvolution calculations habitually employed to determine longitudinal magnetic field from stellar 
 Stokes-V spectra \citep{Donati_1997} can be applied. Models for unperturbed electronic states \citep{Semenov2016,Western2017} can give a useful starting point, but need refining.  
 To initiate this study, some laser excitation spectra were recorded (at UNB) in static magnetic fields up to 0.6 tesla. 
Figure~\ref{fig:FeHeaZeeman}(a) illustrates a simple Zeeman pattern, with a splitting of 0.61 cm$^{-1}$ (=0.17~$\AA$) between the two $\sigma$ profiles of a P line with unresolved $\Lambda$-doubling.
 The individual M$_J$ components are unresolved at Doppler resolution ($\approx$ 0.01$\AA$) with our discharge source.  The Zeeman pattern shown in Fig.~\ref{fig:FeHeaZeeman}(b)  for the Q$_2$(3.5)
  line is more complex, with well-resolved  $\Lambda$-doubling in zero-field conditions. 
 At 0.56 tesla, the partially resolved M$_J$ structure of the Q$_2$(3.5) line extends over 0.7 \AA. Contributions from the two $\Lambda$-components overlap. 
 The e~$^6\Pi$-~a$^6\Delta$  system clearly has potential for magnetic field diagnostics, and we are pursuing this study in the laboratory.

Analysis yields effective Land\'e factors intermediate between Hund's case (a) and Hund's case (b) limiting values. 
    For P$_1$(4.5),  g$^{\prime}_J$ =1.393  and g$^{\prime\prime}_J$ =  1.344 ; for Q$_2$(3.5), we find g$'_{Je}$ =0.906,  g$^{\prime}_{Jf}$=0.887  and g$^{\prime\prime}_{Je,f}$ =  1.312. The statistical uncertainties were 0.001.

 \subsection{FeH quartet line lists (infrared)}
\begin{figure}[!h]
 \centering\includegraphics[scale=0.25]{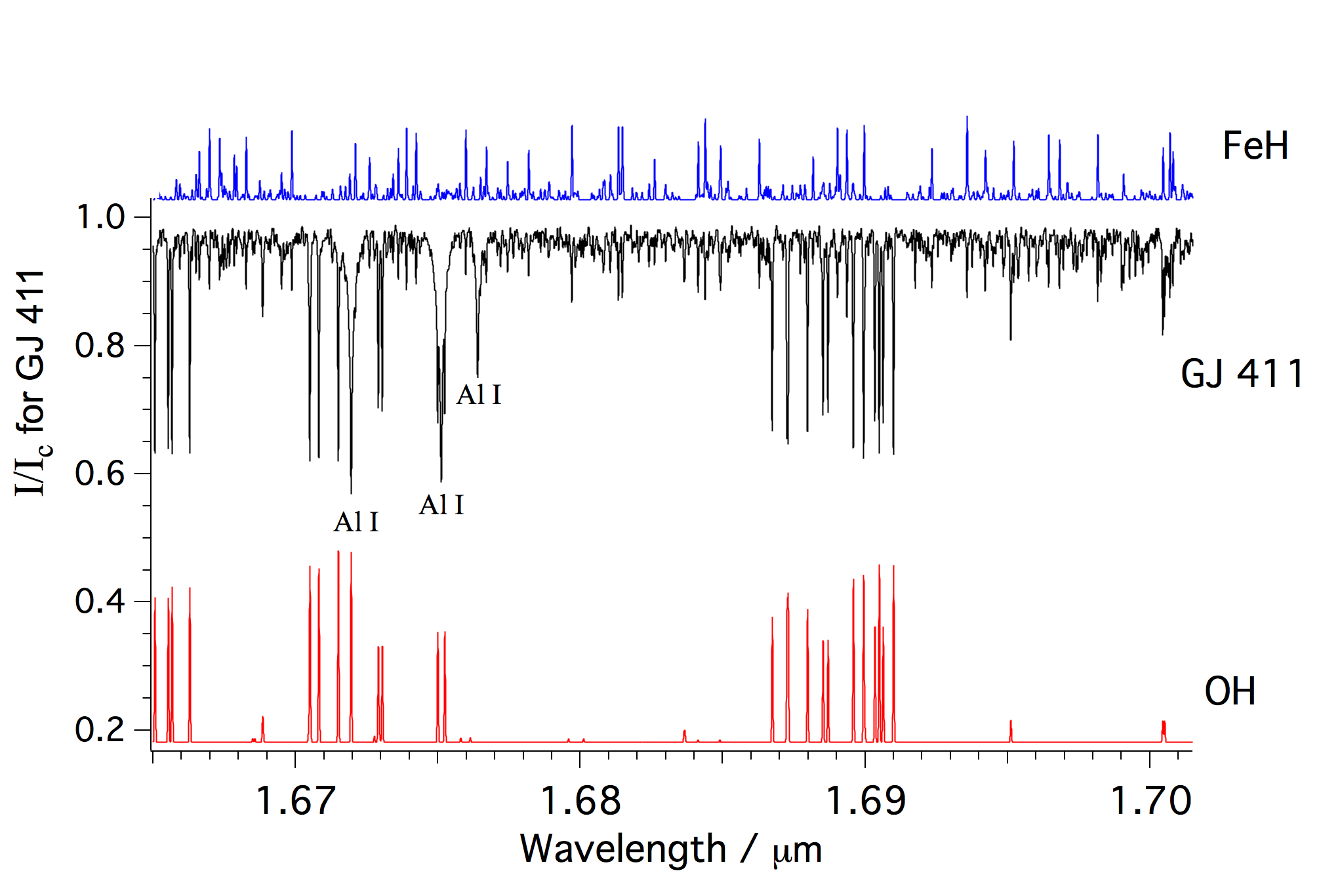}
\caption{ Spectrum of  GJ~411 recorded on SPIRou (corrected for RV) with reference spectra of OH (Hitran database; \citealt{Gordon2022}) and FeH E-A bands in blue taken from the NSO spectrum.
  OH features dominate, but many matches with FeH are obvious. The I/I$_c$ scale refers to GJ~411, corrected for baseline continuum. The scales for FeH and OH are arbitrary.}
\label{fig:GJ411Hitran}
\end{figure}

 The F ${^4\Delta}$ - X $^4\Delta$  system of FeH around 1 $\upmu$m has been extensively studied, and is well established as a diagnostic for conditions in cool stellar objects \citep{Wallace_1998,Dulick2000,Shulyak2010,Reiners_Basri2007}.   We focus here on the E~${^4\Pi}$ - A $^4\Pi$  system in the H band,  and the  E ${^4\Pi}$ -  X $^4\Delta$ system in the J band, 
 as they are ideally suited to  detection with SPIRou, and for analysis of stellar spectra.  
An extensive high-temperature line list for the E~${^4\Pi}$ - A~$^4\Pi$  system around 1.6 $\upmu$m and  for the E~${^4\Pi}$ -  X~$^4\Delta$  near 1.3 $\upmu$m, (see Fig.~\ref{fig:FeHAbInitio}), was published by \cite{Balfour2004}, 
 following their analysis of a  thermal emission spectrum of FeH produced in a King furnace (T~2400 $\deg$ C), recorded at the NSO at Kitt Peak (NSO archive catalogue 1983/01/19\#2).   
 \citet{Wallace2001} searched for these transitions in stellar spectra recorded in the 1990s, but reported 
  `A cross correlation of E-A line positions from the King furnace spectrum with an archival  high-resolution 4 m FTS spectrum of the M2 dwarf GJ~411 
   shows a clean peak at the stellar velocity even though the S/N ratio is not high enough to allow detection of individual lines'.
 The situation is much improved now, with the sensitivity and resolution offered by SPIRou.  
 The SPIRou spectrum from GJ~411 shows very clear matches with the NSO spectrum in the E-A region (see Fig.~\ref{fig:GJ411Hitran}), and a very similar plot is obtained for AD~Leo (GJ~388). This rehabilitates FeH in the 1.3--1.6 micron region for stellar diagnostics. \cite{Souto2020} also mentioned this, and added a caveat to the effect that  the reliability of FeH as a  temperature indicator is limited by the present precision of $g$ factors.
 
\begin{figure}[!h]
 \centering\includegraphics[scale=0.22]{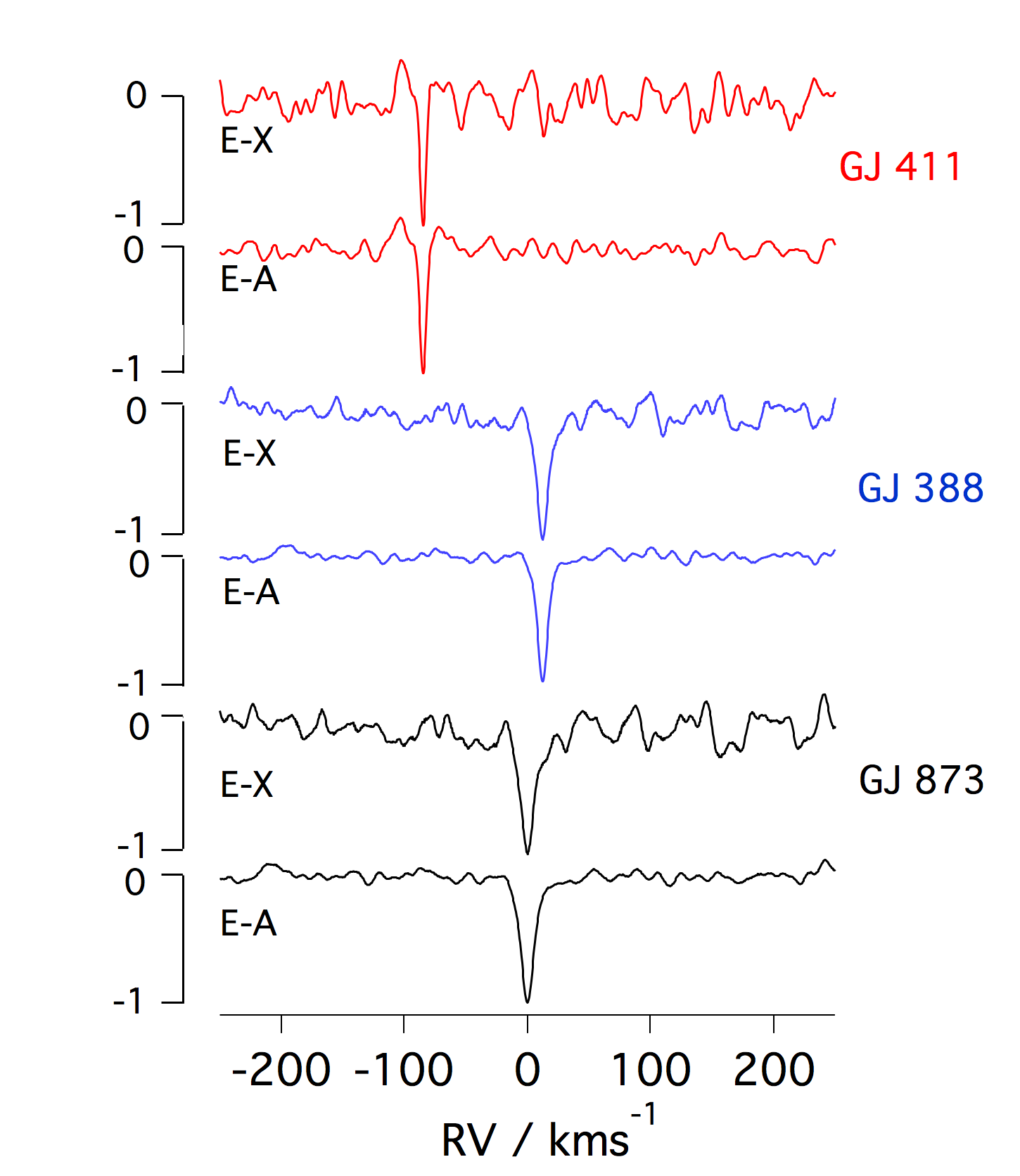}
\caption{CC curves for the Stokes-I spectrum of stars  GJ~411, GJ~388, and GJ~873 with line positions and intensities extracted from the NSO spectrum of FeH (Kitt Peak archive).
2094 lines are listed in the  E-X system 6320--7400 cm$^{-1}$ (1.35--1.58 $\upmu$m), and 1613 in the  E-A system 5200-6322 cm$^{-1}$ (1.58--1.92$\upmu$m). 
Laboratory data were weighted by relative intensities in both calculations.}
\label{fig:correlationsFeHEXEA}
\end{figure}

Figure~\ref{fig:correlationsFeHEXEA}  illustrates the outcome of  CC calculations to determine RVs for the stars 
GJ~411, GJ~388 and  GJ~873 from spectra recorded with SPIRou (including barycentric Earth RV corrections)
using  the E-X and E-A line lists.   The threefold difference in the S/N in the correlation functions derived from these two infrared bands  partly reflects the transition dipole moments 
for the two electronic systems: the E-X bands were actually weaker in the NSO spectrum than E-A, despite a higher population density in the electronic ground state. 
Magnetically inactive GJ~411 gives the sharpest correlation function.  For magnetically active GJ~388, the E-A system at 1.6 $\upmu$m gives RV=12.24(1) km\,s$^{-1}$ and the E-X mask 12.22(1) km\,s$^{-1}$  with a slightly broader correlation peak.  The values are compatible with literature values.
Both masks  perform less well for GJ~873, although the RV determination of  -0.08(1) km\,s$^{-1}$ remains compatible with the values of $0.28\pm0.13$ and $0.36\pm0.13~{\rm km\;s^{-1}}$ derived in \cite{Morin2008}. Degraded performance could equally be the consequence
 of more than one species contributing to the stellar spectrum.  Both the E-X and E-A masks are currently less than optimal because we used a zero-field mask for lack of Land\'e factors when both GJ~388 and  GJ~873 are known to be magnetically active  (see Table~\ref{table:stars}). 
 
Approximate Land\'e factors for the E-A system, estimated from spectropolarimetric studies of sunspots \citep{Asensio2004}, suggest that neither electronic state follows the limiting Hund's case coupling schemes, 
but no high-resolution laboratory data measurements are yet available. Our attempts to remedy this were unsuccessful, probably because the chemistry producing FeH from iron pentacarbonyl 
and hydrogen atoms yields mostly ground state products. In fact, we failed to detect E-A transitions even in zero field conditions in laser excitation, using a low-power diode laser and InGaAs detector.
 Zeeman studies were therefore out of the question. Future work will probably aim to use pump-dump laser experiments via stronger transitions to obtain this information.
 
 \subsection{CrH A-X system at 862 nm}
\begin{figure}[h!]
 \centering\includegraphics[scale=0.2]{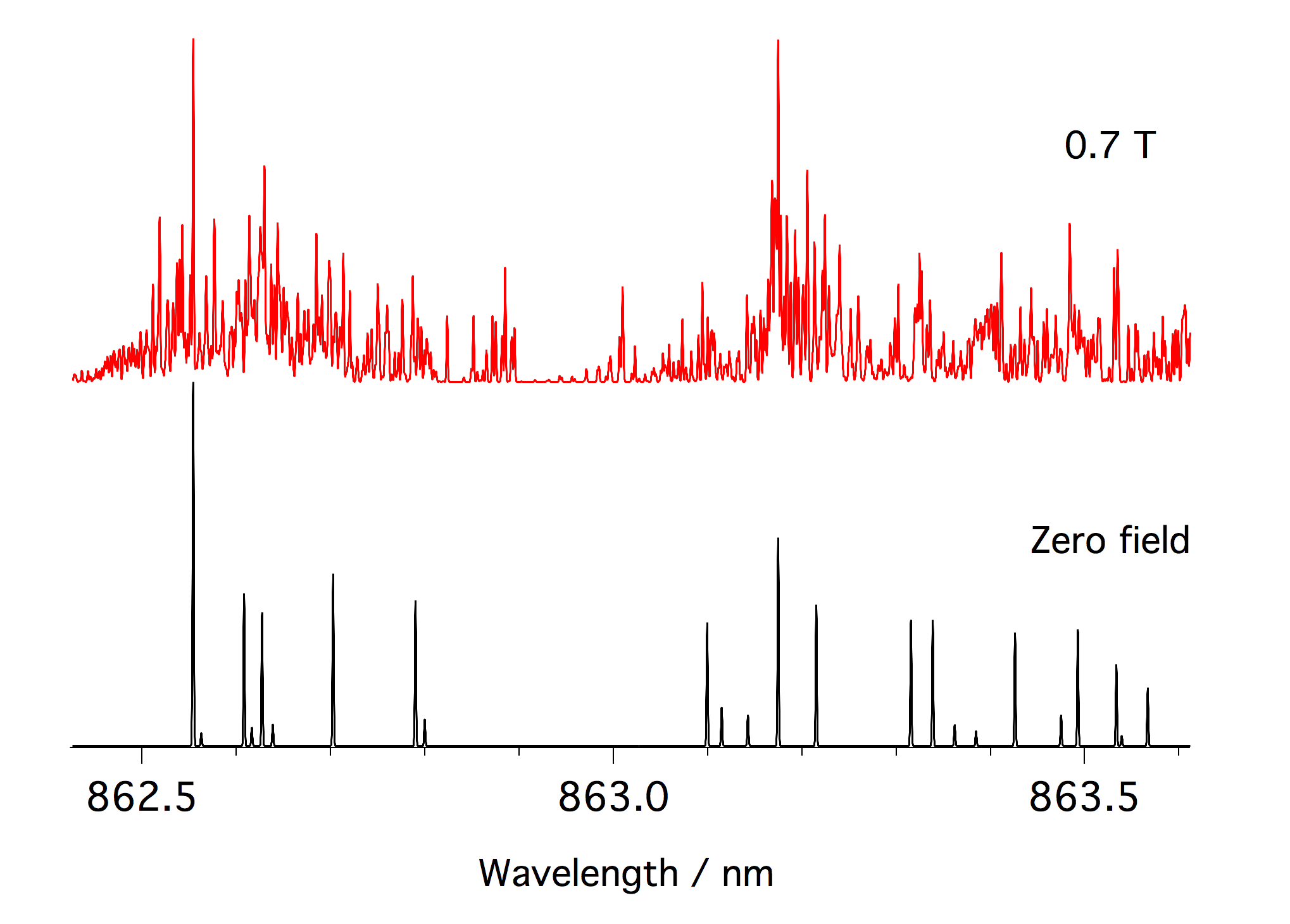}
\caption{ Calculated R lines near the CrH A-X (0-0) band origin. The black line is at zero field. The red line is a simulation at  0.7 tesla, ($\Delta$M$_J$ 
= $\pm$1) transitions only, with the vertical scale multiplied by 5.}
\label{fig:CrH-B0-and-Zeeman}
\end{figure}

 CrH has been adopted as a diagnostic tool in the context of L-type brown dwarf stars \citep{Kuzmychov2013,Kuzmychov2017} and even exoplanetary atmospheres (\cite{Braam2021}), 
 but less for the atmospheres of M-type stars (\cite{Pavlenko2014}).  It would be reasonable to expect CrH signatures to appear alongside, or even overlapping with, FeH features, 
 between 750 nm and 1 $\upmu$m. 
We included Paschen-Back effects in CrH to improve spectral masks and then used them with the CC technique to test for the presence of CrH in some red dwarf stars.
 The calculated mask changes enormously with magnetic field, as illustrated in Fig.~\ref{fig:CrH-B0-and-Zeeman} where the upper trace shows a 
 simulation with $\Delta$M$_J$  = $\pm$1)  transitions only, corresponding to $\perp$ linear polarisation in the laboratory with respect to magnetic field orientation.

\begin{figure}[h!]
 \centering\includegraphics[scale=0.25]{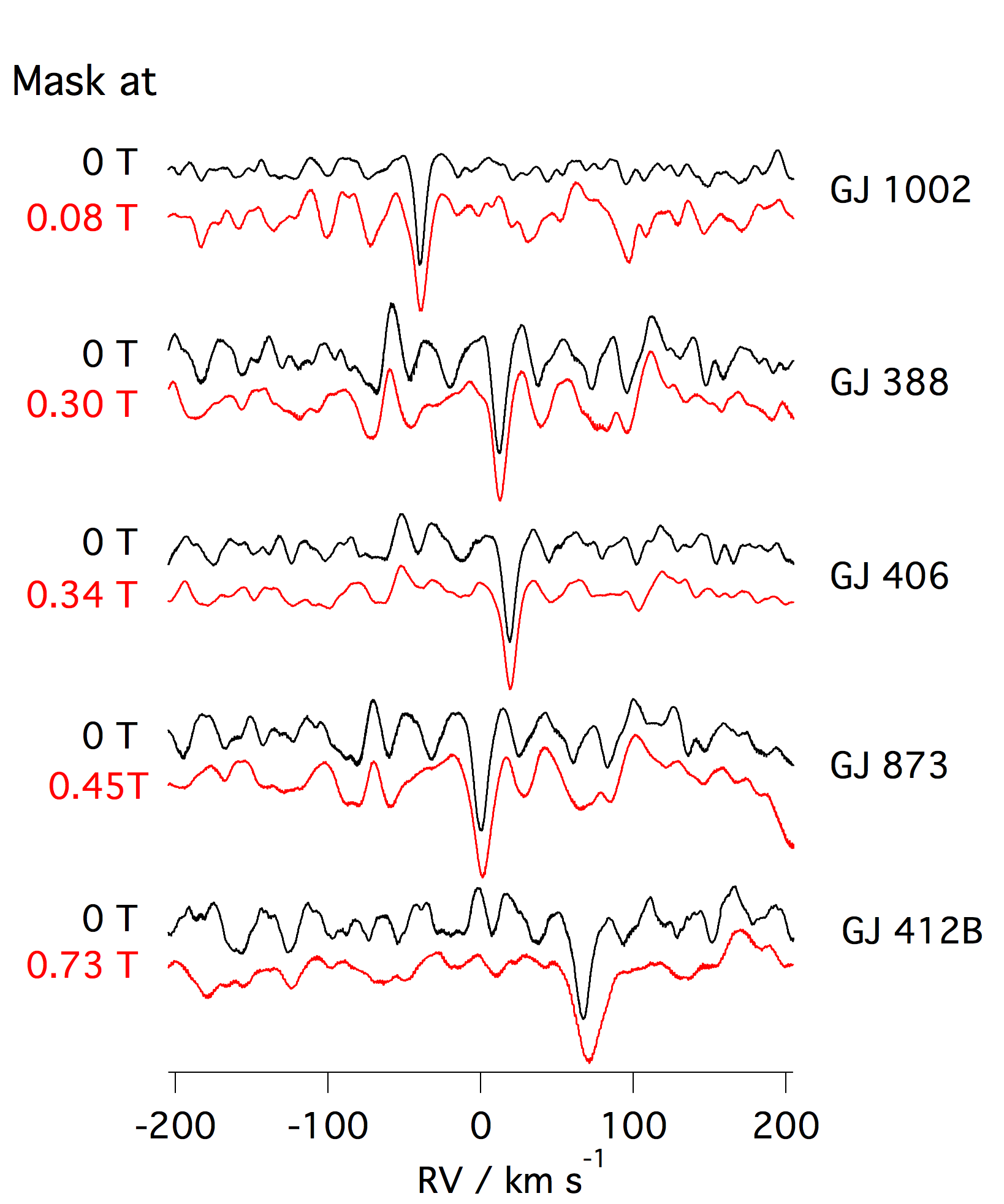}
\caption{ CCs (normalised to unity)  between spectra of increasingly magnetically active stars and CrH masks.  
A zero-field mask was used to compute the results shown in black. The strongly magnetically active stars give less noisy correlation plots when the
Paschen-Back effect is included (shown in red). The correlation peak positions are sensitive to <B>. This is clearest for GJ~412B.}
\label{fig:correlationsCrHAX}
\end{figure}

We performed CC function calculations, taking masks first from the usual zero-field atlas from \cite{Burrows2002} and then from a series of predictions generated with parameters from our PGOPHER fits, taking the  <B> values cited in Table \ref{table:RVresults}.
 The PGOPHER predictions between 11502 and 11610 cm$^{-1}$, or 862 -- 869 nm, were restricted to upper state levels v=0, N$^\prime~\le$~15, and calculated for $\Delta$M$_{J}$ = $\pm$1 transitions at chosen magnetic field strength.  
  The outcome is shown in Fig.~\ref{fig:correlationsCrHAX}, with the correlation function with the zero-field atlas shown above the result for non-zero <B> in each case.  The reference atlas performed very well for GJ~1002, which is 
 known to be a weakly active star with low magnetic flux.  The peak at RV -39.71(2) km\,s$^{-1}$ matches determinations from atomic data.  The zero-field atlas actually gives a flatter baseline and narrower correlation peak than the  PGOPHER mask at 0.08 T, from which the RV value is  -39.31(2) km\,s$^{-1}$.

 The correlation functions taken from field-free masks are particularly unconvincing for GJ~873, T = 3341 K, <B> = 0.453 tesla, and GJ~412B, T = 3000 K, <B> = 0.73 tesla 
 although the telescope spectra are of comparable quality.  
 The baselines of the correlation  functions calculated for the reported  <B> values (red lines) are somewhat smoother in Fig.~\ref{fig:correlationsCrHAX}.  
 However, it is also clear that the sharpness of the correlation function decreases with magnetic field strength.
Several factors contribute. One is that the spectrum becomes quasi-continuous under the combined effect of Zeeman/Paschen-Back splittings. A second is that large `$v$\,sin$i$' 
 contributions from (rapid)  stellar rotation broaden individual line profiles, and a third  is that an average field strength in the direction of observation is assumed, 
 ignoring any stellar magnetic field complexity. This will also  blur  the spectrum.  
 
Having found that the CC functions with CrH data yielded plausible RV values, though performing less well for the stars with low RVs, 
we tried to use a similar approach to see if we could also
 determine average magnetic field strengths from CC peaks, using the S/N in the correlation peak as criterion for optimisation. We took GJ~412B as a test case. 
 Masks were calculated for <B> = 0 to 0.8 tesla, in steps of 0.1 T. The S/N was highest (7) for the mask calculated at 0.6 T, giving a RV of 69.12(1)  km\,s$^{-1}$.  
 At 0.7 T, the RV determined from the CC function was 69.78(2) km\,s$^{-1}$.  
 This implies that the A-X system of CrH is not necessarily the best diagnostic for magnetic field because of the density of lines, unless reliable RV information is available
 from other observations.  Given a strong constraint on RV, the CC functions allow us to retrieve average magnetic field strengths to $\pm$ ~0.1 T for  GJ~412B and GJ~873  
spectra.  Laboratory work on CrH is still in progress, and we will discuss the use of CrH for M-dwarf magnetometry in a subsequent paper.

\begin{figure}[!h]
 \centering\includegraphics[scale=0.3]{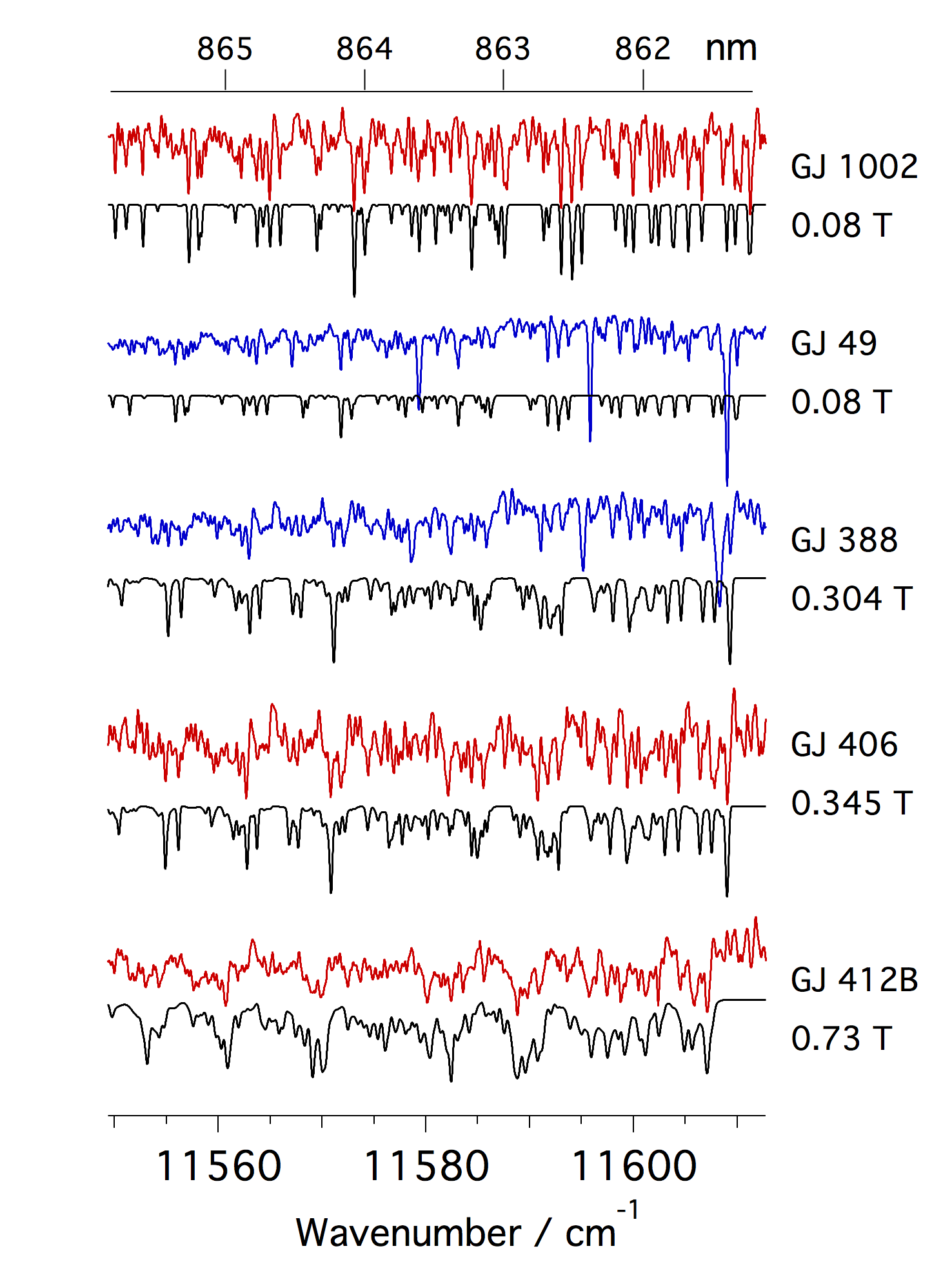}
\caption{Comparison between models of part of the A-X (0,0) band of CrH 865.56 -- 862.57 nm, with spectra (in colour)  from the Narval (GJ~49 and GJ~388) and ESPaDOnS (GJ~1002, GJ~406, and GJ~412B)  instruments.
Blue curves are for stars whose temperatures are $>$ 3000 K. The upper scale indicates $\uplambda_{\textnormal{vac}}$ in nm.}
\label{fig:Espadons_vs_PGOPHER}
\end{figure}

Figure~\ref{fig:Espadons_vs_PGOPHER} compares the spectra from ESPaDOnS close to the origin band of the A-X system of CrH with simulations from PGOPHER.  
The simulations took convolutions of line lists with profiles taking into account stellar rotation, limb-darkening, micro- and macro- turbulence, instrumental resolution and Doppler widths. The stellar parameters come from the literature (see Table 2). 
CrH features are not difficult to recognise in this R-head region for the cooler stars in our selection. There are some matches for the warmest star in our selection, GJ~49 with T = 3805 K. 
Even though the visual match for GJ~388 is less convincing, the corresponding correlation function  in Fig. \ref{fig:correlationsCrHAX} suggests CrH is present.  
FeH persists at higher temperatures: the FeH Wing-Ford bands near 1 $\upmu$m are more prominent in these spectra.

 \subsection{NiH: Visible systems}
\begin{figure}
 \centering\includegraphics[scale=0.25]{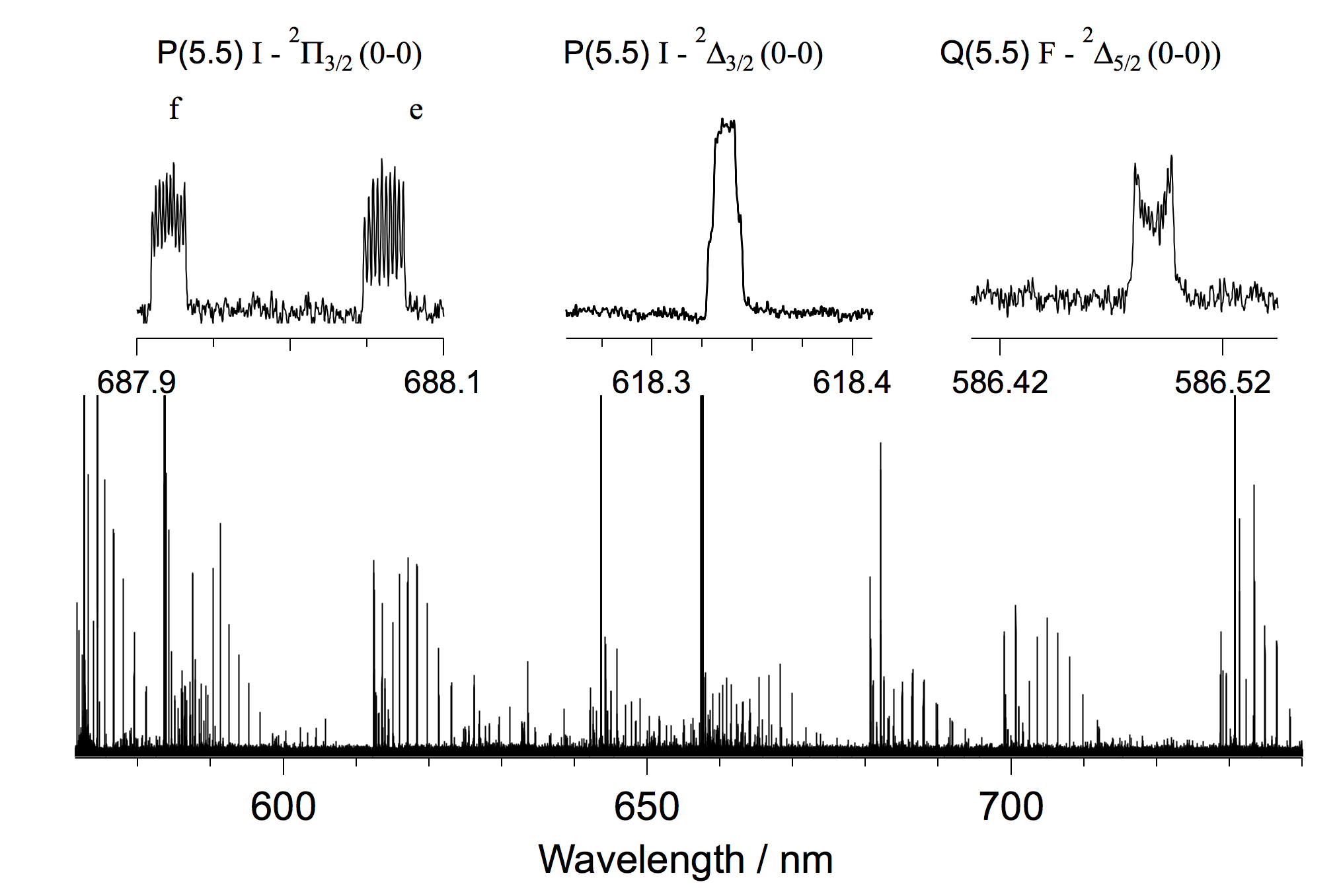}
\caption{ Laboratory fluorescence spectrum of NiH recorded at 0.7 T. The upper graphs show three individual lines  in  4.5 cm$^{-1}$ windows to illustrate the difference 
in Zeeman response at a given rotational quantum number in three electronic systems. Note that the Zeeman patterns differ even for the two $\Lambda$-doublet components of the P(5.5) line, in the top-left plot.}
\label{fig:NiHZeeman}
\end{figure}

\begin{figure}
\centering\includegraphics[scale=0.28]{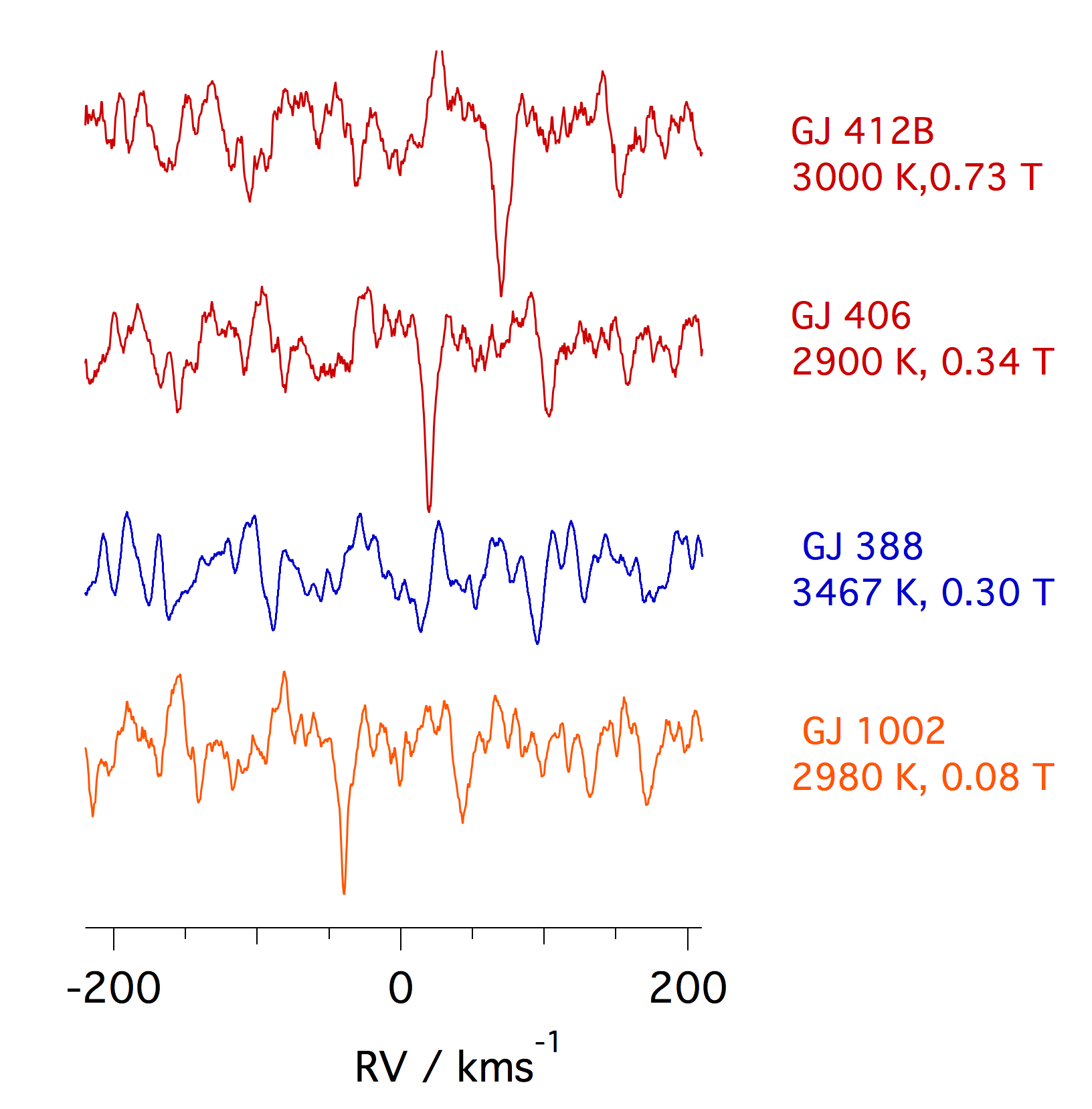}
\caption{CC functions (normalised to unity) for Stokes-I spectra of four sample M stars using the mask from the zero-field NiH laboratory work of \cite{Vallon2009} and \cite{Havalyova2021}. The S/N is poor, but the RVs of the strongest peaks match values derived from atomic data.  NiH is not detected in GJ~388, one of the warmest stars in our sample.}
\label{fig:NiH-CCFs}
\end{figure}

 Some 50 years ago, a few features in sunspot spectra were attributed to NiH \citep{Lambert_1971,Wohl_1971}, but this has remained controversial. 
 As far as we know, there have been no reports of NiH signatures in the optical spectrum of M-type stars.  Figure \ref{fig:NiHZeeman} shows that features from several electronic systems of NiH cover the optical spectrum, without forming distinctive band-heads seen for CrH and FeH. 
 There is no obvious region where NiH features are likely to dominate a stellar spectrum. However, working with this mask taken directly from dispersed fluorescence 
 spectra recorded in Lyon \citep{Vallon2009,Ross2012,Harker2013}, we now have some evidence from CC peaks displayed in Fig.~\ref{fig:NiH-CCFs} that NiH does contribute to the Stokes-I spectra of M stars. 
Only three of the M-type stars in our selection gave correlation peaks with the full NiH mask  cited in Sect. 3.2 (2167 lines), which
 covers a spectral window 9970--19000 cm$^{-1}$ (570--740 nm), and assumes that all three lowest-lying electronic states are thermally populated in the stellar source.  
 Fuxrther tests showed there was no need to include the full laboratory spectrum in CC calculations. Only transitions between 13000 and 17500 cm$^{-1}$  
 originating from the v=0 and 1 levels of the electronic ground state  X$_1\;^2\Delta_{5/2}, $ and from v=0 of  X$_2\;^2\Delta_{3/2}$,   
  actually contributed to the correlation peak.  The optimised line selection is included in supplementary data files. 
 Figure~\ref{fig:NiH-CCFs} shows the resulting CC curves using the reduced reference mask.  
Neither GJ~388 (illustrated in Fig.~\ref{fig:NiH-CCFs}) nor GJ~49, the warmest stars considered here, showed  any evidence of correlation with the NiH line list, but this (modest) increase in temperature does not explain our observation.
Gaussian fits  to the CC functions of Fig.~\ref{fig:NiH-CCFs} gave RVs very close to the values given in the literature (see Table~\ref{table:RVresults}). 
Testing with a mask of NiD lines gave no correlation at all, also suggesting that  the small correlation peaks are not due to chance coincidences in line positions. 
We conclude that NiH is present  in these mid M-dwarf atmospheres, although we cannot say which of the electronic
bands make significant contributions.  It is unsatisfactory in that we cannot identify individual lines of NiH, but these (often widely spaced) features 
overlap with more closely spaced lines in strong bands of TiO across the visible and near-infrared spectrum, and make only minor contributions to overall opacity.
\begin{table*} [ht]
\caption{RVs determined from molecular data, and the parameters (taken from references quoted in Table~\ref{table:stars}) used for spectral simulations.
Magnetic field effects were included for CrH simulations only. Data sources:  $^a $\cite{Lafarga2020}, $^b$\cite{Cristofari2022}; \cite{Cristofari2023}, $^c$\cite{Morin2010},
$^d$\cite{Fouque2023},  $^e $\cite{Donati2008}, $^f $\cite{Kochukhov2021}, $^g$\cite{Reiners2022} Table B1.}

\label{table:RVresults}
 \centering                           
 
\begin{tabular}{l c c c c c c c}
 
\hline
RV (km\,s$^{-1}$) from                          &GJ~411         & GJ~1002                 &    GJ~49    &      GJ~406 &    GJ~388 &      GJ~873    &     GJ ~412B \\ 
                                                & Lalande 21185                 &                                 &                    & CN Leo      &    AD Leo     &     EV Lac     &      WX UMa    \\  
\hline
CrH A-X (0.86 $\upmu$m)          &                &      -39.31(2)              & -4.96(1)        &  +19.25(1)      & +12.58(1)           &  +1.18(1)       &  +70.81(2)  \\ 
FeH E-A (1.6 $\upmu$m)          &  -84.89(1)   &                        &                         &                     & +12.24(1)       &  -0.08(1)     &              \\
FeH E-X (1.3 $\upmu$m)          &  -84.92(2) &                                  &                         &                      & +12.22(1)      &  -0.08(3)     &              \\
FeH (vis)                               &                  &    -40.01(3)                & -6.4(1)       & +19.45(3)             &                       &                    &+ 69.1(2) \\
NiH (vis)                               &                  &    -39.7(1)                 & (no CCF)      &  +19.8(1)     & (no CCF)      &                    & +70.0(1)  \\
\hline
Reference RV (km\,s )$^a $      & -85.01(2)   &  -40.12(2)     & -6.30(2)                 &  +19.28(2)    &+12.29(2)      &+0.35(3)      &  +69.16(2)  \\
                                                
\hline  
\multicolumn {4}{l}{Parameters for simulations}  &   &  &      \\

 T (K)                                  &  3601(51)$^a$& 2980(33)$^b $            & 3805         & 2899(31)$^b $  & 3467(31)$^b $   & 3341(31)$^b $   & 3000  \\  
v sin $i$ (km\,s$^{-1}$)                        &    1.5~[<2]$^a$       & <0.1$^{b,d} $    & 1$^e$        &    2(1)$^b $      & 3(1)$^b $            &  3(1))$^b $              & 5$^c$  \\
Limb Darkening $\epsilon$               &    0.3                & 0.5              & 0.5                 & 0.5           & 0.3                           & 0.3                     & 0.5 \\
Micro-turbulence $\xi$ (km\,s$^{-1}$)        & 0.85      &  0.85                &  0.85           &  0.85                 & 0.85                  &                               & 0.85\\
Macro-turbulence $\zeta$ (km\,s$^{-1}$)   &1.35         & 1.35          & 1.35            & 1.35          & 2.3                   & 2.3                   &1.35\\
Instrument resolving power              &  70000                & 65000                 & 65000           & 65000                 & 70000                         & 70000                   & 65000\\
<B> (tesla)                                     &  <0.02$^g$   & 0.08$^f $       & 0.08(2)$^f $  & 0.345(20)$^b$ & 0.303(6)$^b $         & 0.453(7)$^b $       & 0.73(3)$^f $   \\
\hline
\end{tabular}
\end{table*}

The overview of RV determinations presented in Table~\ref{table:RVresults} shows that RVs are for the most part determined to within 0.5 km\,s$^{-1}$  of the reference values taken from
Lafarga's compilation. The 1~$\sigma$  statistical uncertainties quoted in units-of-last-digit, obtained by fitting the clean part of the correlation peak (away from baseline noise) to a Gaussian 
function, may be  deceptively small. It is not unusual for absolute RV values to be sensitive to choice of mask and method. \cite{Carmona2023} discuss this in detail for AD~Leo, deriving  RV~=~12.695(5) km\,s$^{-1}$ with SPIRou data and an infrared mask, and RV~=~12.50(2) km\,s$^{-1}$ when processing optical domain data with SOPHIE (Spectrographe pour l'Observation des PH\'enom\`enes des
Int\'erieurs stellaires et des Exoplan\`etes); in both instances, dozens of measurements taken over many months were averaged. 
Both values are considered  reliable despite disagreement within their quoted uncertainties, because the critical quantity is the relative precision  of measurement. Regular or random fluctuations in RV can be detected and interpreted with respect to a mean value, and it is the variations around the mean that are important for planet detection.

\section{Conclusion}

We investigated a small selection of early-to-mid M-type main sequence stars of known small-scale  fields, <B>,  up to 0.7 tesla, working with spectra recorded with the ESPaDOnS and SPIRou instruments at the Canada France Hawaii Telescope and Narval at the Bernard Lyot Telescope.  We used existing 
optical and near-infrared laboratory data for three molecular species, FeH, CrH, and NiH, adding new laboratory measurements for CrH to establish 
reference line lists. These lists (available for download from the CDS) were then used to compute CC functions for our target stars in order to: 
confirm or refute the detectability of these transition metal MHs in the high-resolution spectra of cool dwarf stars; and 
assess their potential for precise velocimetry and magnetometry. 
The line lists are intended to be used either as a binary mask or with intensity patterns predicted from Boltzmann populations and theoretical transition strengths, as
implemented by \cite{Burrows2002}. 
 They should be a useful supplement to established atomic
 reference lines, particularly when band structures are at least partly resolved, because they have the advantages of rarely being saturated and having less extensive spectral wings. 
 
We illustrate the dramatic change in the absorption spectrum expected to occur when the magnetic response of CrH in its  $^6\Sigma^{+}$ states is taken into account, working with an effective 
Hamiltonian model to reproduce field-free spin-rotation splittings in the excited state, for low rotational levels.  Even with this restricted database, 
 we show that RVs and even magnetic field strengths can be predicted from molecular data alone.  
Optimising  the wavelength range to be included in correlation calculations reduces computation time and limits  the influence of chance coincidences with other species 
present. Above 869.4 nm, the CrH A-X 0-0 band overlaps with the 1-0  band of the F-X system of FeH, so it may be appropriate to truncate the CrH mask to focus on diagnostics from chromium hydride alone.
In doing this, we find evidence for CrH in two of the warmer M stars in our selection, GJ~388 (AD~Leo) and GJ 49.  This was unexpected since, according to \cite{Allard2013},  CrH is expected to appear clearly only in cooler atmospheres, when metal monoxides have condensed and contribute less to the gas phase spectrum.
 Our work shows that the 860 nm band of CrH, already identified as a valuable tool for L-type brown dwarf magnetometry (\cite{Kuzmychov2017}), can also potentially help characterise the magnetic fields of mid-to-late M-dwarf stars.  Stokes-V measurements are likely to be particularly important in this context because of the spectral density arising from spin-rotation couplings 
 in this optical system. Our simulations suggest that <B> values of the order of 0.08 T could be determined readily from high-resolution Stokes-V spectra,
  where Stokes-I features would remain unresolved and difficult to analyse from line profiles.
  
  Our study, using  CC techniques in conjunction 
  with well-chosen electronic bands, has focused on showing that transition metal hydrides can be detected in a small selection of cool stars.\ Therefore, adding these new transitions to existing masks should help improve statistics for detection, perhaps even in exoplanets!  The possible signatures of CrH in a hot Jupiter, WASP-31b, reported by \citet{Braam2021} were confirmed this year by \citet{Flagg2023} and provide a huge incentive to pursue laboratory investigations.

\section {Acknowledgements} {This work was supported by the French Programme National de Physique Stellaire of CNRS/INSU, (project LABSPIR / AO INSU AA 2022).  We also acknowledge funding from NSERC Canada, and we thank the SPIRou Legacy Survey consortium for sharing data from the SPIRou instrument. 
JFD  and AK acknowledge funding from the European Research Council (ERC) under the H2020 research \& innovation programme via
 grant agreements \#740651 NewWorlds (JFD ), and \# 716155 SACCRED (AK).  SB, PF, AC, JF-D,JM,   acknowledge funding from the French ANR under contract number ANR\-18\-CE31\-0019 (SPlaSH), and from the Investissements d'Avenir programme (ANR-15-IDEX-02) via the `Origin of Life' project of the Grenoble-Alpes University.

This work is based on observations obtained at the Bernard Lyot (TBL) and Canada France Hawaii (CFHT) telescopes. 
The TBL is operated by the Institut National des Sciences de l'Univers of the Centre National de la Recherche Scientifique of
France (INSU/CNRS). The CFHT is operated by the CNRC (Canada), INSU/CNRS (France) and the University of Hawaii. 
The authors wish to recognise and acknowledge the very significant cultural role and reverence that the summit of Maunakea has always had within the indigenous Hawaiian community. 
We are most fortunate to have the opportunity to conduct observations from this mountain. 
This work also benefited from the SIMBAD CDS database at URL \url{http://simbad.u-strasbg.fr/simbad}, the NASA ADS system at URL \url{https://ui.adsabs.harvard.edu}, the PolarBase database at URL \url{http://polarbase.irap.omp.eu/}, and the ExoMol database at URL \url{https://www.exomol.com}.}

\bibliography{CCF_biblio_Abbr}

\bibliographystyle{aa.bst}

\end{document}